\documentclass[prl,twocolumn,times,showpacs,superscriptaddress]{revtex4-1}
\usepackage{amsmath, amssymb, braket, dsfont, times, tensor}
\usepackage[mathscr]{euscript}
\usepackage{graphicx}
\usepackage{subfigure}		
\usepackage{epstopdf, times}
\usepackage[breaklinks]{hyperref}

\usepackage[usenames,dvipsnames]{color}			

\newcommand{\bse}{{\boldsymbol{e}}}

\newcommand{\cs}{{C_6}}
\newcommand{\bea}{\begin{eqnarray}}
\newcommand{\eea}{\end{eqnarray}}
\newcommand{\imth}{\hspace{1pt}\mathrm{i}\hspace{1pt}}

\begin{document}
\title{Space group symmetry fractionalization in a chiral kagome Heisenberg antiferromagnet}
\author{Michael P. Zaletel}
\affiliation{Station Q, Microsoft Research, Santa Barbara, California 93106-6105, USA}
\author{Zhenyue Zhu}
\affiliation{Department of Physics and Astronomy, University of California, Irvine,  Irvine, CA 92697, USA}
\author{Yuan-Ming Lu}
\affiliation{Department of Physics, The Ohio State University, Columbus, OH 43210, USA}
\author{Ashvin Vishwanath}
\affiliation{Department of Physics, University of California, Berkeley, California 94720, USA}
\author{Steven R. White}
\affiliation{Department of Physics and Astronomy, University of California, Irvine,  Irvine, CA 92697, USA}
\date{\today}							
\begin{abstract}
The anyonic excitations of a spin-liquid can feature fractional quantum numbers under space group symmetries. Detecting these fractional  quantum numbers, which are analogs of the fractional charge of Laughlin quasiparticles, may prove easier than the direct observation of anyonic braiding and statistics.
Motivated by the recent numerical discovery of spin-liquid phases in the kagome Heisenberg antiferromagnet, we theoretically predict the pattern of space group symmetry fractionalization in the kagome lattice chiral spin liquid.
We provide a method to detect these fractional quantum numbers in finite-size numerics which is simple to implement in DMRG.
Applying these developments to the chiral spin liquid phase of a kagome Heisenberg model, we find perfect agreement between our theoretical prediction and  numerical observations.

\end{abstract}

\maketitle


Two-dimensional  quantum spin liquids are distinguished by emergent excitations,  `spinons,' which carry an  $S=1/2$ moment,  in striking contrast to all local excitations (e.g. magnons) which carry integer spin.
Like the fractional charge of the Laughlin quasiparticles,\cite{Laughlin1983} their spin is an example of `symmetry fractionalization:'  symmetries can act on topological excitations in a way which is  forbidden for the local excitations.
Wen proposed that  in  addition to charge and spin, the quantum numbers of space group symmetries, like translation, could also become fractional. \cite{Wen2002, WenSN2003} Subsequent work revealed  a zoo  of distinct gapped spin-liquid phases distinguished by their fractional space group quantum numbers. \cite{Wen2002, WenSN2003, Kitaev2006, Wang2006, Lu2011, Essin2013, Mesaros2013, Barkeshli2014}

It is important to understand the pattern of symmetry fractionalization in a spin liquid since it provides one of the few potential experimental probes of fractionalized spin liquid physics.
For example, space group fractionalization has spectroscopic signatures,\cite{Wen2002, WenSN2003, EssinHermele2014} and determines the nearby ordered phases that are connected to the spin liquid via continuous phase transitions.\cite{Sachdev1992,Wang2006,Lu2011a,Lu2014,Lu2015}
Great theoretical progress has been made in the classification of space group symmetry fractionalization,\cite{Essin2013, Barkeshli2014} though fractionalized space group quantum numbers have yet to be detected in a Heisenberg spin model.\cite{Song2015}

Here we report the direct detection of space group symmetry fractionalization in a Heisenberg antiferromagnet on the kagome lattice.
Recently, several works have discovered that  introducing chiral symmetry breaking terms \cite{Greiter2014, Bauer2014}  or further-neighbor exchange interactions \cite{HeChenYan2014, gong2014emergent,Wietek2015}  can stabilize a chiral spin liquid (CSL).
Proposed by Kalmeyer and Laughlin, the CSL is the magnetic analog of the $\nu = \frac{1}{2}$ bosonic quantum Hall effect, with a robust spin-carrying gapless edge protected by its chiral central charge $c = 1$.\cite{Kalmeyer1987, WWZ1989, Greiter2009}
The CSL contains a single type of anyonic excitation, the  $S=1/2$ spinon `$s$,' which has semionic statistics with itself.
In close analogy to the Laughlin flux-threading argument, when $2\pi$-flux of the $S^z$ spin rotation (or any other axis) is thread through the system, the flux nucleates a spinon $s$.
Since $s$ carries $S^z = \pm \tfrac{1}{2}$ itself, the flux insertion has induced spin, which is the famous spin-Hall response $\sigma_{xy}^{\textrm{spin}} = \pm \tfrac{1}{2}$.
The sign of the response depends on  the parity-breaking chirality.

	Previous studies have confirmed SO(3) symmetry fractionalization in the kagome CSL, which can be detected from the fractional spin-Hall response\cite{gong2014emergent}.
In this work, we theoretically predict the pattern of space group symmetry fractionalization in the kagome CSL, and conduct numerical experiments using large-scale cylinder DMRG to detect this pattern in the  $J_1$-$J_2$-$J_3$ Heisenberg model.
space group symmetries are not simple to probe in `snake' DMRG, since the chosen 1D ordering of the sites breaks the spatial symmetries.
We introduce a technique, the classical product state (CPS) trick, for detecting space group symmetry fractionalization.
The CPS trick also drastically simplifies the measurement of the topological S and T matrices,\cite{Zhang2012} which previously required Monte Carlo sampling  as expensive as the DMRG itself.\cite{CincioVidal}
Using both finite and infinite DMRG, and several cylinder geometries, we find perfect agreement with theoretical predictions.
The methods introduced here are applicable to many other spin liquid models.

\emph{Theory of symmetry fractionalization in a CSL.}
\begin{figure}[h]
\includegraphics[width=0.6\linewidth]{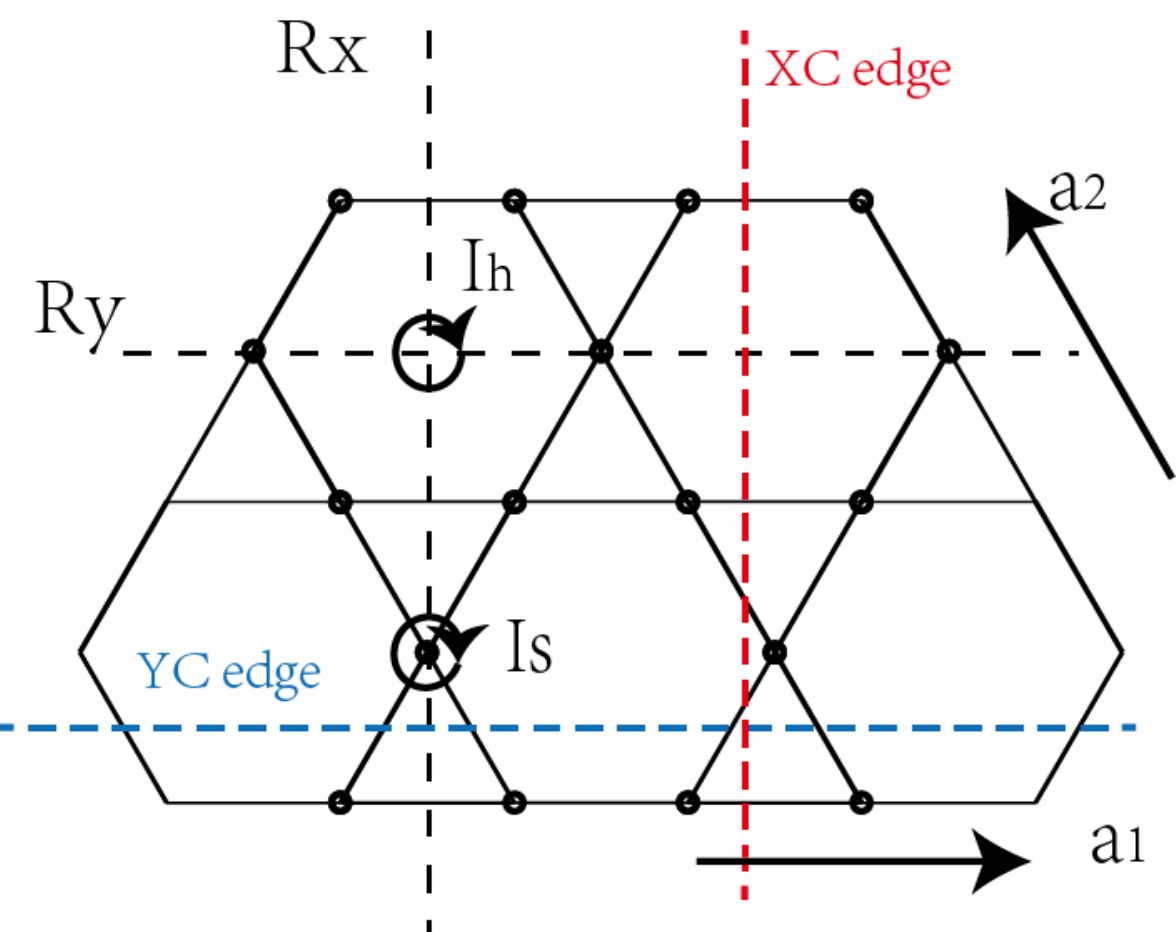}
\caption{Symmetry operations and two typical edges in numerical studies of the kagome chiral spin liquid.
\label{fig:symmetry group}}
\end{figure}
In addition to  SO(3) rotations of spins, the kagome model has a number of space group symmetries illustrated in Fig. \ref{fig:symmetry group}.
$T_{1,2}$ denote translations along Bravais vectors $\mathbf{a}_1, \mathbf{a}_2$, and $C_{6}$ is a hexagon-centered $\pi/3$ rotation.
In particular there are two inequivalent inversion operations: hexagon-centered $I_h=(C_6)^3$, and site-centered $I_s=T_1 I_h$.
Due to the chiral order parameter of the CSL, both reflection symmetry and time-reversal are spontaneously broken.
However, their combination is preserved, so we define  \emph{anti}-unitary reflections $R_x, R_y$, whose orientation with respect to the Bravais vectors is illustrated in Fig. \ref{fig:symmetry group}.
The space group generators satisfy the algebraic conditions summarized in the left column of Tab.~\ref{tab:PSG}, where $\bse$ represents the identity element.

The symmetry fractionalization\cite{Wen2002, Essin2013, Barkeshli2014} of the CSL is encoded in how symmetry operations act on individual spinons.
For example, when the inversion $I_h$ acts on a spinon, it may acquire a phase `$\pm i$,' which is `fractional' since on local objects $I_h = \pm 1$.
There are a number of symmetry-group relations which can be similarly fractionalized when acting on spinons, which we tabulate in the left column of Tab.~\ref{tab:PSG}.
In each case, there is a group relation that should produce the identity (like $I_h^2 = \bse$) which instead produces a phase.
There is an important constraint on the phase: since a pair of spinons annihilates to the vacuum, $s\times s=\mathds{1}$, which can't be fractionalized, the phases are $\mathbb{Z}_2$-valued, $\pm1$.
The phase factors associated with the last two algebraic identities in Tab.~\ref{tab:PSG} are not well-defined and can be fixed as $+1$ by a proper gauge choice. These $\mathbb{Z}_2$-valued phase factors are the topological invariants labeling a CSL on kagome lattice.

\begin{table}[tb]
\centering
\begin{tabular} {|c|c|c|}
\hline\hline
Algebra&SET invariants&Measurements\\
\hline
$T_1T_2T_1^{-1}T_2^{-1}=\bse$&$  -1$&$e^{\imth (P_s-P_{\mathds{1}})}$\\
\hline
$(\cs)^6=(I_h)^2=\bse$&$-1$&$Q_s(I_h)/Q_{\mathds{1}}(I_h)$\\
\hline
$(R_x)^2=\bse$&$-1$&${R_x}$-SPT on YC8\\
\hline
$(R_y)^2=\bse$&$-1$&${R_y}$-SPT on XC8\\
\hline
$R_xT_1R_x^{-1}T_1=\bse$&$+1$&${R_xT_1}$-SPT on YC8\\
\hline
$R_xT_2R_x^{-1}T_2^{-1}T_1=\bse$&$-1$&${R_yT_y}$-SPT on XC8\\
\hline
$\cs T_1\cs^{-1}T_2^{-1}=\bse$& $+1$ (gauge fixing)&N/A\\
\hline
$\cs T_2\cs^{-1}T_2^{-1}T_1=\bse$&$+1$ (gauge fixing)&N/A\\
\hline\hline
\end{tabular}
\caption{Group relations and predicted spinon fractionalization of the kagome CSL.}
\label{tab:PSG}
\end{table}

We now derive the fractionalization pattern of the CSL and a set of concrete measurements to detect it.
Each of the 6 independent SET invariants can be numerically measured from the degenerate ground states of a long (or infinite) cylinder.\cite{Zaletel2015}

Consider first the relation $(I^2_h)_s = -1$.
Measuring  such a phase seems a contradiction, since when $I_h$  acts on a finite number of spins it must give $\pm 1$ by its very definition.
The key insight is that  rather than trying to act with $I_h$ on a single spinon, we create a pair of spinons related by $I_h$, and measure the global $I_h$ quantum number of the pair.
Strictly speaking, we are interested in the quantum number relative to the vacuum.
If $-1$, it is as if $I_h \cdot s= \pm i \cdot s$ when acting on each individually, which indicates fractionalization.
The robustness of this procedure was argued in Ref.\cite{Zaletel2015}.

\begin{figure}[h]
\includegraphics[width=1\linewidth]{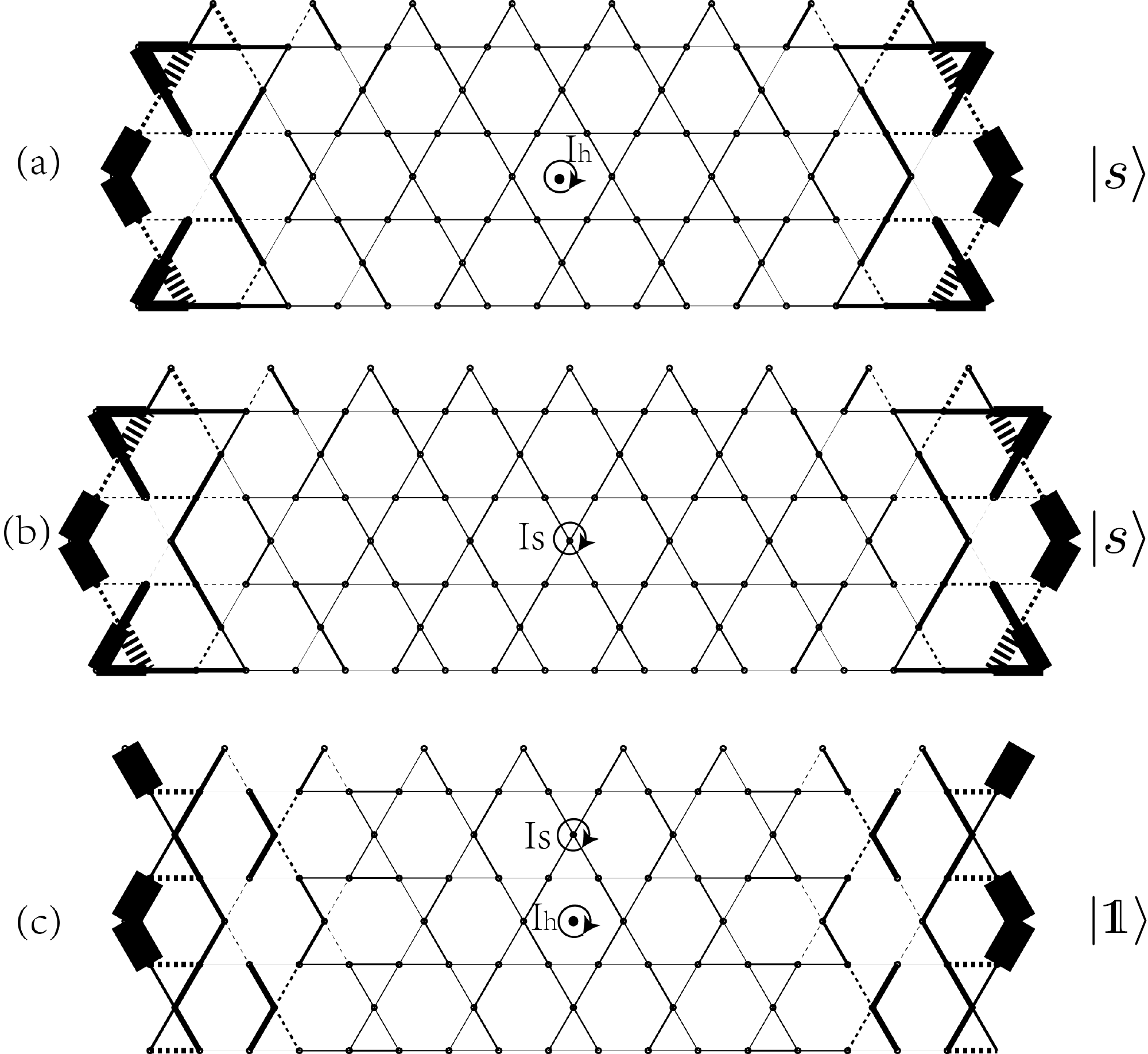}
\caption{Finite DMRG geometries used for XC8. Geometries (a) and (b) have an odd number of edge spins, trapping a spinon, from which we compute $Q_s(I_h), Q_s(I_s)$. From (c), we compute both  $Q_{\mathds{1}}(I_h)$ and $Q_{\mathds{1}}(I_s)$. The results are tabulated Fig.~\ref{fig:csl_hist}.
A similar set of geometries is used for YC8.
\label{fig:geometry}}
\end{figure}
In practice, it is not necessary to nucleate and manipulate a spinon pair.
Instead, we make use of topological ground state degeneracy.
Like a torus, an infinitely long cylinder has a two-fold ground state degeneracy.
A useful basis choice are the minimally entangled states (MES),\cite{Kitaev2006a, Dong2008,Zhang2012} which are labeled by the two topological sectors: $\{ \ket{\mathds{1}}, \ket{s} \}$.
Given the ground state $\ket{\mathds{1}}$, the ground state $\ket{s}$ is obtained by nucleating a pair of spinons and separating them out to infinity.
If we instead use a finite cylinder (Fig.~\ref{fig:geometry}) the  pair eventually encounters the boundaries; since we must leave one at each edge, there is a energy splitting between $\ket{\mathds{1}}, \ket{s}$, but this is purely a boundary effect.
The ratio of $I_h$ quantum numbers $Q_{\mathds{1}/s}(I_h)$ in these two states reveals the fractionalization of the spinon:
\begin{align}
\frac{Q_s(I_h)}{Q_{\mathds{1}}(I_h) } = (I_h)^2_s.
\end{align}

We now prove that $(I_h^2)_s=-1$ for a CSL, so long as SO(3) is preserved, by using the flux-insertion trick introduced in Ref. \cite{Hermele2015, Qi2015a}.
The spinon sector $\ket{s}$ can be obtained from the vaccum state $\ket{\mathds{1}}$ by adiabatically threading $S^z$-flux $\phi$ through the cylinder, i.e., by twisting the boundary conditions.
Due to the spin-Hall response, when $\phi = 2\pi$,  $\Delta S^z = \pm \frac{1}{2}$ of spin has been transferred from one end of the cylinder; this should be interpreted precisely as the spinon sector $\ket{s}$, since the spinons brought to the edge bring with them a magnetic moment.
Clearly the $S^z$ flux $\phi$ will be inverted ($\phi \to -\phi$) by either inversion $I_h$ or $\pi$ spin rotation $e^{\imth\pi S^x}$, but (with a proper choice of branch cut) it will remain invariant under their combination $e^{\imth\pi S^x}I_h$. Therefore we can track the eigenvalue of $e^{\imth\pi S^x}I_h$ throughout the flux insertion process, which must remain unchanged:
\bea
\frac{Q_s(e^{\imth\pi S^x}I_h)}{Q_{\mathds{1}}(e^{\imth\pi S^x}I_h)}=1=\big[(e^{\imth\pi S^x}I_h)^2\big]_s 
\eea
As noticed in Ref. \cite{Qi2015a}, to be compatible with the continuous SO(3) spin rotational symmetry, a plaquette-centered inversion operation must commute with all spin rotations when acting on the semionic spinons. Therefore we have
\bea
\big[(e^{\imth\pi S^x}I_h)^2\big]_s=(e^{\imth2\pi S^x})_s\cdot(I_h^2)_s=1 
\eea
Since each semion carries spin-$1/2$, $(e^{\imth2\pi S^x})_s = -1$, and we have proved that $(I_h^2)_s=-1$.

	The spinon may experience an effective $\pi$-flux per unit cell, $(T_1T_2T_1^{-1}T_2^{-1})_s = - 1$.
To detect it, consider a cylinder periodic in $\vec a_2$ with length $L_1$.
In the $s$ sector there is a spinon trapped at the edge, so by growing the length $L_1 \to L_1 + 1$ we  effectively act on a single spinon by $T_1$.
Measuring the resulting change in the momentum $Q_{s}(T_2)$ around the cylinder then reveals $(T_1T_2T_1^{-1}T_2^{-1})_s$.
Specifically, we measure the difference in the two sectors' `momentum per unit length:'
\begin{align}
\frac{e^{i Q_{s}(T_2) } }{e^{i Q_{\mathds{1}}(T_2) }}= a \cdot [  (T_1 T_2 T_1^{-1}T_2^{-1})_s ]^{L_1}.
\end{align}

To derive this phase, recall the $s$ sector is obtained by threading $2\pi$ $S^z$ flux through the cylinder.
Since the kagome magnet has half-integral spin per unit cell, Oshikawa's argument,\cite{Oshikawa2000} which generalizes the Lieb-Schultz-Mattis theorem, will apply.
The argument predicts that threading flux increases the momentum around the cylinder by $\pi$ for each unit length along $\vec a_1$ direction, i.e., it changes the momentum per unit length.
So $(T_1 T_2 T_1^{-1}T_2^{-1})_s = -1$.

The site-centered inversion $I_s$ is related to a hexagon-centered inversion $I_h$ by a lattice translation. After fixing the last two lines of Tab.~\ref{tab:PSG} by a proper gauge choice, we have
\bea
\frac{Q_s(I_s)}{Q_{\mathds{1}}(I_s) } = (I_s^2)_s=(T_1T_2T_1^{-1}T_2^{-1})_s\cdot(I_h^2)_s=+1\notag
\eea
Thus it is sufficient to measure $I_h$ and $I_s$ quantum numbers.

Now let's turn to line 3-6 in Tab.~\ref{tab:PSG} which are related to anti-unitary reflection symmetries. Consider a YC2n cylinder on which anti-unitary reflection $R_x$ does not exchange the two edges, i.e., it acts like an ``on-site time reversal'' symmetry.
As shown in Ref.\cite{ZaletelSPT2014, Huang2014, Zaletel2015}, each ground state sector can be regarded as a gapped 1d spin chain with a symmetry protected topological (SPT)\cite{PollmannTurnerBO2010, FidkowskiKitaev2011, ChenGuWen2011} invariant.
In the presence of on-site antiunitary symmetry $G=\mathbb{Z}_2^{R_x}$, their SPT invariants take value of $H^2\left[ \mathbb{Z}_2^{R_x},\textrm{U}(1)^{G} \right ] = \mathbb{Z}_2=\{+1,-1\}$.
The trivial phase ($+1$) typically has gapped symmetric edges, while the nontrivial SPT phase ($-1$) supports $R_x$-protected zero modes on the edge similar to a Haldane chain.
Therefore the $\mathbb{Z}_2$-valued SET invariant $(R_x^2)_s$ is given by the ratio of $R_x$-SPT invariants for spinon and vacuum sectors.

Here we prove that $(R_x^2)_s=-1$ must hold for any CSL whose spinons carry half-integer spin each, using the flux insertion trick again.\cite{Hermele2015, Qi2015a} Since $S^z$ flux $\phi$ adiabatically inserted through the cylinder is invariant under the combined operation of spin rotation $e^{\imth\pi S^x}$ and anti-unitary reflection $R_x$,\footnote{Note that the location of the twist boundary condition requires one to choose a `defect line' whose location  breaks $R_x$; thus the symmetry $e^{\imth\pi S^x} R_x$ implicitly includes a `gauge-transformation' $e^{i \phi S^z}$ on a subset of the spins in order to bring the defect line back to the same position. This transformation doesn't change any of the group relations.}
the vacuum and semion sector must share the same SPT invariant associated with anti-unitary $\mathbb{Z}_2$ symmetry $e^{\imth\pi S^x}R_x$ and hence
\bea
\big[(e^{\imth\pi S^x}R_x)^2\big]_s=(e^{\imth2\pi S^x})_s\cdot(R_x^2)_s=1\notag
\eea
As a result we have $R_x^2=-1$ for spin-$1/2$ semions. The same argument leads to $(R_xT_1)^2=(R_y)^2=(R_yT_y)^2=-1$ for semionic spinons where $T_y=T_1^{-1}T_2^2$.
Now straightforward algebra can show that all SET invariants summarized in Tab. \ref{tab:PSG} are fixed.

\emph{Absolute quantum numbers.}
Thus far we have  predicted only the relative quantum numbers between topological sectors. Under certain assumptions,  the absolute quantum numbers can be predicted as well.
Consider a cylinder whose ground state has no free moments, i.e., $\langle \mathbf{S}_i \rangle = 0$ on all sites.
Note that for energetic reasons, this might not always be the case, but the couplings could in principle be tuned to ensure it.
If the ground state remains moment-free when adding a pair of spins, one at each edge, the introduced moments must be `screened' by pair creation of  spin-1/2 excitations, i.e.,  spinons, which sit precisely on the additional sites.
Note that the spin-1/2 character of the added sites was essential, otherwise a \emph{local} excitation could screen the new moment.
As the entire cylinder can be built up this way, the lattice behaves like a `crystal' of semionic spinons.

We now assume that the global quantum number $I_h$ of the geometry can be computed by taking this picture literally and applying $I_h$ to each pair of spinons (i.e., sites).
This assumption is certainly true within the parton construction, \cite{Zaletel2015, Qi2015} but we contend it holds more generally.
On the one hand, we have already determined that $(I_h)^2_s = -1$, so we find
\bea
&Q(I_h)=(-1)^{\#~\text{of}~I_h\text{-pairs}}
\eea
where $I_h\text{-pairs}$ denotes pairs of $I_h$-related sites in the lattice.
This prediction has a nontrivial dependence on the cylinder type, since on  XC8, the central column contains a single pair of sites, while on YC8, it contains two.
However, we would like to propose a more  universal way of computing the result which does not depend on the results of the preceding section.
Under  $I_h$ the spinon on each lattice site exchanges with its inversion counterpart, and meanwhile each spinon also rotates by $\pi$ around itself. Due to
the semionic statistics of spinons, each counter-clockwise exchange will contribute a phase $e^{\imth\pi/2}$ per inversion-related pair, while counter-clockwise self rotation by $\pi$ leads to phase $e^{\imth\pi/4}$ per spinon.
During this exchange, the trajectory of each pair always encloses an even number of the other semions, so there is no further phase.
Therefore the total phase obtained in this process is again
\bea
&Q(I_h)=e^{\imth\frac\pi2\frac{N_s}2}\cdot e^{\imth\frac\pi4N_s}=(-1)^{\frac{N_s}2}=(-1)^{\#~\text{of}~I_h\text{-pairs}}
\eea
where $N_s$ denotes the total number of lattice sites.

When computing $I_s$, two contributions differ.
First, two of the sites are left invariant, so they do not acquire a phase.
Second, the exchange of all the remaining pairs encloses \emph{one} of these stationary sites, acquiring an extra mutual statistics $(-1)^{N_s/2 - 1}$.
Together,
\bea
&Q(I_s)=(-1)^{\#~\text{of}~I_s\text{-pairs}}\cdot (-1)^{\#~\text{of}~I_s\text{-pairs}}\equiv+1.
\eea

Since this computation depends only on braiding and statistics, it is interesting to speculate how it extends to a $\mathbb{Z}_2$ spin liquid, where there are different possible symmetry fractionalization patterns which lead to different global quantum numbers.
In the $\mathbb{Z}_2$ spin liquid there are two distinct types of spinons (bosonic and fermionic) which could sit on the sites, and there is a spinless `vison' excitation which could be placed in various plaquettes.
Thus, unlike the CSL, there are multiple different `anyon crystals'  consistent with the location of the $S=1/2$ moments, depending on the spatial arrangements of visons.
Intriguingly, this suggests that space group fractionalization is encoded in a particular anyon crystal.

\emph{Detecting symmetry fractionalization: the CPS trick.}
If one had access to the wave function as a dense vector---as in exact
diagonalization---it would be trivial to compute the needed global quantum
numbers.  However, DMRG maps the cylinder to a 1D chain and then compresses the
wave function as a matrix product state (MPS). A symmetry operation $\hat{U}$ mapping sites
to sites can in general be written as a (long) product of nearest neighbor swap operators.
One can then, in principle, calculate the
symmetry overlap $Q_U = \bra{\Psi} \hat{U} \ket{\Psi}$ by sequentially applying the swaps to $\ket{\Psi}$, recompressing
the MPS in the process, and
then doing a final MPS-MPS overlap calculation. In practice, the intermediate states produced when applying the
swaps can have more entanglement than the ground state itself, requiring bigger bond dimensions, and this method
is rather slow and unsatisfactory.

\begin{figure}[t!]
\includegraphics[width=0.8\linewidth]{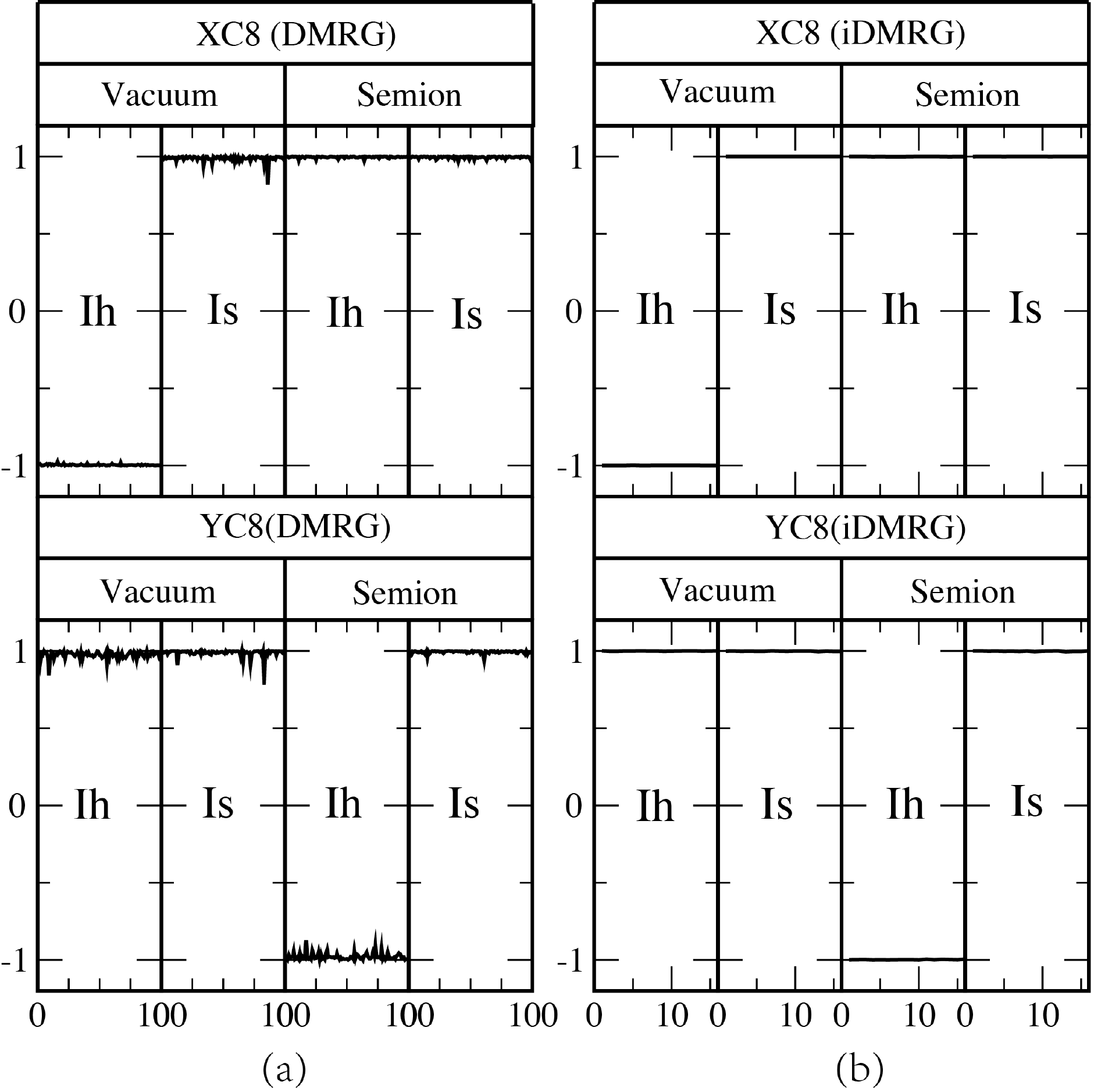}
\caption{Measured reflection quantum numbers $Q_{\mathds{1} / s} ( I_{h/s})$ from a large set of CPS overlaps.
Results were obtained using both finite and infinite DMRG, on the YC8 and XC8 cylinders.
\label{fig:csl_hist}}
\end{figure}

A much better approach involves sampling $\ket{\Psi}$: choosing random product states (classical product states, or CPS) according to the probability distribution $|\Psi|^2$. \cite{StoudenmireWhite2010}
In contrast to a calculation time of $O(N m^3)$ for a DMRG sweep, where $N$ is the number of sites and $m$ is the bond dimension, finding each independent sampled CPS requires only $O(N m^2)$ operations, so  thousands of CPS can be obtained in the time of a single sweep.
Expanding the wavefunction in terms of a complete set of CPS $\{\sigma\}$,  $\ket{\Psi} = \sum_\sigma a_\sigma \ket{\sigma}$,  if $U \ket{\Psi} = Q_U \ket{\Psi}$, then
\begin{align}
a_{U \sigma} = Q_U a_\sigma,
\end{align}
where $U\sigma$ is another CPS trivially obtained from $\sigma$.
Thus a single pair of CPS amplitudes,  $a_\sigma$ and $a_{U \sigma}$, are enough to determine $Q_U$.
Given $\sigma$, the calculation time for $a_\sigma = \braket{\sigma |  \Psi}$ is also $O(N m^2)$.
In practice, since $\ket{\Psi}$ is approximate,  we obtain a distribution $Q_U(\sigma)$, but its mean is
$\bra{\Psi} \hat{U} \ket{\Psi}$, and if $\ket{\Psi}$ is accurate the distribution is sharply peaked at the
correct value.
Thus, we typically sample hundreds or thousands of CPS, plotting the distribution, which gives both the
symmetry eigenvalue and a sense of how certain it is.
A typical example is shown in Fig.~\ref{fig:csl_hist}.  There is very little
noise, indicating $\ket{\Psi}$ is very nearly symmetric.

Similarly, while in principle it is known how to measure  1D-SPT invariants in infinite  DMRG \cite{PollmannTurner}, the existing algorithm is cumbersome in the present case where the reflection permutes the DMRG snake.
We find  a CPS trick can also be used to measure the 1D-SPT invariants of an infinite cylinder, as detailed in the SI.

The CPS trick can be extended to other measurements that require space group operations.
For example, in order to compute the topological  $S$ and $T$ matrices
\cite{Zhang2012} given the degenerate ground states of a torus $\{ \ket{a} \}$,
one must compute overlaps of the form $\bra{b} \hat{R}_{\theta} \ket{a}$ where
$R_\theta$ is a  rotation of the torus.  In the context of MPS this computation
is more difficult than DMRG, but in the SI we show the result can be trivially
computed from a handful of CPS overlaps.
It is quite remarkable that braiding ($S$) and statistics ($T$) are encoded
entirely in the overlaps of the ground state manifold with a handful of CPS.

\emph{Results.}
We study the CSL phase at $J_1 = 1.0$ , $J_2=J_3 = 0.5 $
using complex wavefunctions.
We have computed the inversion quantum numbers on XC8 and YC8 cylinders, using both finite\cite{Steve1992}  and infinite DMRG.\cite{McCulloch2008, Kjall2013}
In finite DMRG, the topological sector is changed by removing sites from the end of the cylinder,\cite{YanHuseWhite2011} as illustrated  in Fig.~\ref{fig:geometry}.
In infinite DMRG, the two sectors appear as a ground state degeneracy.\cite{CincioVidal, Zaletel2013}
In all cases, the relative quantum numbers $Q_s / Q_{\mathds{1}}$ are in agreement with predictions, as summarized in Fig.~\ref{fig:csl_hist}.
Furthermore, recall that for YC8 geometries in the vacuum, we predicted $(-1)^{\#~\text{of}~I_h\text{-pairs}} = 1$, while for XC8 $(-1)^{\#~\text{of}~I_h\text{-pairs}} = -1$.
This difference is reflected in the observed absolute quantum numbers.

\begin{figure}[t!]
\includegraphics[width=0.8\linewidth]{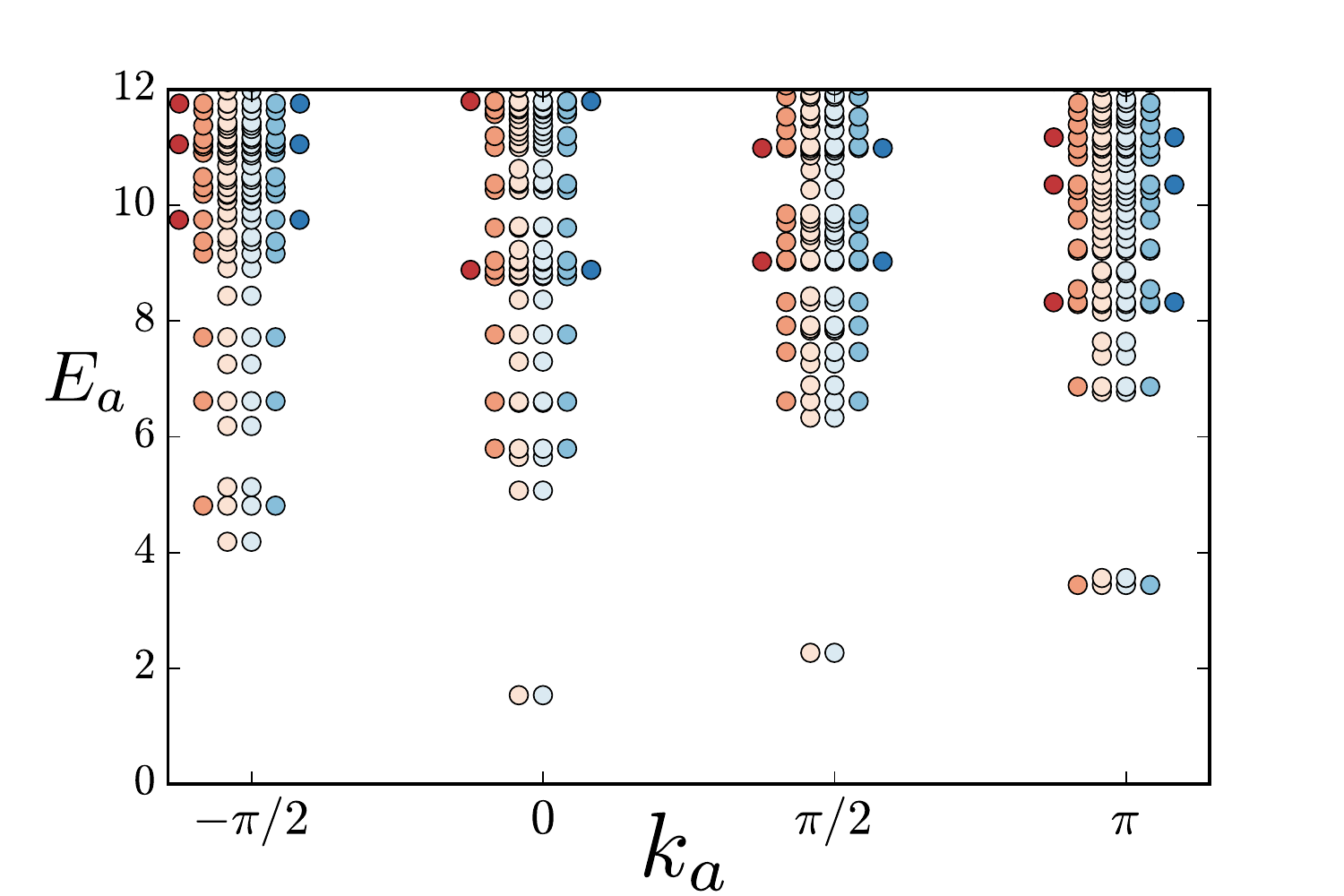}
\caption{Entanglement spectrum $\{E_a\} = -\log(\rho_L)$ of the YC8 $s$-sector, plotted against the momentum $k_a$ around the cylinder of the Schmidt states. The lowest pair is a `Kramers doublet' under the anti-unitary $R_x$, indicating $(R_x^2)_s = -1$. Since the two levels occur at the same $T_1$ momentum, we find $(R_x T_1)^2 = - 1$ on the pair as well, as expected from $(R_x T_1)^2_s = -1$.
\label{fig:ent_spec}}
\end{figure}

To measure $(R_x^2)_s$, we measure the $\mathbb{Z}_2$ 1D-SPT invariant associated with $R_x$ using iDMRG.
The details of this measurement are discussed in the SI, but the result, $(R_x^2)_s = -1$, is apparent from the entanglement spectrum of the $\ket{s}$ sector,  shown in Fig.~\ref{fig:ent_spec}.
The spectrum has a two-fold degeneracy, which was verified to transform as a Kramers doublet under $R_x$; the $\ket{\mathds{1}}$ sector, in contrast, does not.
Note that both levels occur at the same momentum around the cylinder; this implies $T_1$ acts trivially on the levels, so the pair is also a Kramers doublet under $R_x T_1$, implying $(R_x T_1)^2_s = -1$. Similar agreement for $R_y$ is found on XC8.

	In conclusion, we have shown that the CSL state has a unique but non-trivial pattern of space group symmetry fractionalization, and detected this pattern in a microscopic Heisenberg kagome model.
In addition, we have elucidated a general framework for probing crystal symmetries with DMRG,  which can be applied, for example, to other recently discovered spin liquid phases.
	
\begin{acknowledgments}
We thank D Huse for many discussions and his ongoing contributions to related work.
MPZ acknowledges helpful conversations with L Cincio and M Cheng.
YML is supported by the startup funds at Ohio State University.
AV is supported by the Templeton Foundation and by a Simons Investigator grant.
SRW and ZZ acknowledge support from the NSF under grant DMR-1505406 and from the Simons Foundation through the Many Electron collaboration.

\end{acknowledgments}

\bibliography{bibs}

\clearpage
\appendix
\section{The CPS trick for computing the $S$ and $T$ matrices.}
\label{app:CPSrep}
To obtain the topological $S$ and $T$ matrices, it is sufficient to calculate the action of a rotation $\hat{R}^{\theta}$ on the MES basis of a torus,
\begin{align}
R^{\theta}_{ab} \equiv \bra{a} \hat{R}^{\theta} \ket{b}
\end{align}
Here we explain how to obtain the matrix $R^{\theta}_{ab}$, or any other symmetry, using CPS overlaps.

In general, let $g \in G$  denote the symmetry group of the Hamiltonian and  $\{ \ket{a} \}$ be a basis for the set of ground states; they could arise either from topological order or spontaneous symmetry breaking. The ground states form a representation of the symmetry group:
\begin{align}
g_{ab} \equiv \bra{a} \hat{g} \ket{b}, \quad  (gh)_{ab} = \sum_c g_{ac} h_{c b}
\end{align}
If ground states are missing, the matrices $g_{ab}$ will \emph{not} form a representation, a good test for a complete basis.

How do we calculate $g$? Let $ \mathcal{C} = \{ \ket{\sigma} \}$ be some states which a) individually break all the symmetries b) have non-zero amplitude in the ground states, and c) are trivial to apply the symmetries to. Random tensor product spin configurations that are well represented in the wave function should do, e.g. $\ket{\sigma} = \ket{\uparrow \downarrow \rightarrow \cdots}$.
We write $\ket{g \sigma} \equiv  \hat{g} \ket{\sigma}$.
By acting with all elements of the symmetry group, we can extend $\mathcal{C}$ to ensure it forms a representation of  $G$.

In any tensor network scheme it is trivial to measure the overlap between these spin configurations and the ground states:
\begin{align}
V_{\sigma, a} \equiv \braket{ \sigma | a}.
\end{align}
$V$ has nice symmetry properties,
\begin{align}
\bra{\sigma}\hat{g} \ket{a} &=  \braket{\sigma | b} g_{b a} =  \braket{ g^{-1} \sigma  | a} \\
\sum_{b} V_{\sigma, b} \, g_{b a} &= V_{g^{-1} \sigma, a}.
\end{align}
Suppose that $V_{\sigma, a} $, when viewed as a rectangular matrix, has a rank equal to the number of $a$. This can always be achieved by adding more spin configuration $\sigma$ to the set $\mathcal{C}$ (along with all symmetry related configurations). Then $V$ has a pseudoinverse, and the representation of $g$ can be extracted from
\begin{align}
g_{b a} = \sum_{\sigma \in \mathcal{C}} V^{-1}_{b, \sigma} V_{g^{-1} \sigma, a}
\end{align}
Again, the sum $\sigma$ does not need to run over a complete set of basis states in the full Hilbert space.
For $N$ ground states $\{ \ket{a} \}$, typically only $N$ of the $\sigma$ configurations will be required to invert $V$.
So we only need $\mathcal{C}$  to have a handful of spin configurations in order to recover $g$.
Since individual overlaps $V$ may have some noise, the stability of the inverse can be improved  by adding more configurations $\sigma$ to the set $\mathcal{C}$.

For comparison, one possible Monte-Carlo scheme (over physical, rather virtual MPS indices)  proceeds by sampling over
\begin{align}
g_{ab} = \bra{a} \hat{g} \ket{b} = \sum_{\sigma \in \mathcal{H}} \braket{a | \sigma} \bra{\sigma}\hat{g}\ket{b} = \sum_{\sigma \in \mathcal{H}} V^\dagger_{a,\sigma } V_{g^{-1} \sigma, b},
\end{align}
where $\mathcal{H}$ is the entire Hilbert space.
Hence simply by using $V^{-1}$ rather than $V^\dagger$, we need only examine a couple configurations.

\section{Efficiently detecting 1D - inversion protected SPTs in snake iDMRG}
\label{app:iSPT}
There are a multitude of possible ways to use CPS tricks to extract space group 1D-SPT order from iDMRG.
The approach we found most convenient to implement proceeds as follows.
In step one, we cut out a segment of the iMPS in order to obtain a set of ansatz wavefunctions  for a \emph{finite} cylinder, $\{ \ket{a b} \}$, where $a, b$ will run over degenerate `edge states.'
In step two, we use the CPS trick to measure how the $\{ \ket{a b} \}$ transform under the symmetries; the structure of  the resulting representation is sufficient to determine the topological indices.

\subsection{Step 1: symmetric finite cylinder ansatz from iDMRG}
	In order to measure symmetry properties with respect to a group $G$, consider a finite cylinder invariant under $G$ with a length $L$ which is several times the correlation length.
This finite cylinder can be viewed as a subset of an infinite cylinder.
We require that  the iDMRG snake is ordered such that this subset of sites corresponds to a \emph{contiguous} set of sites in the 1D MPS ordering.
This is a  restriction on the ordering of the snake - however, it can always be engineered at the end of the simulation via a handful of swap gates.
Under this condition, the two edges of the finite cylinder correspond to the two MPS bonds at the edge of this contiguous block. A set of ansatz wavefunctions for this continguous block are constructed from the data of the MPS in Vidal's canonical form, $\{ \Gamma, s\}$:
\begin{equation}
\ket{a_1, a_{L+1}} \equiv   \sum_{ \{a_i\}, \{p_i\} } \Gamma^{p_1}_{a_1 a_2} s_{a_2} \cdots  s_{a_L} \Gamma^{p_L}_{a_L a_{L+1}}  \ket{p_1 \cdots p_L}.
\end{equation}
The indices $a_i$ run over the bond dimension of the MPS.
In canonical form, the indices $a_i$ label Schmidt states, which are organized into degenerate multiplets which transform into each other under the symmetries.
On the terminating bonds, we restrict $a\equiv a_1, b \equiv a_{L+1}$ to lie in the \emph{lowest} multiplet, which we take to have dimension $d_L, d_R$ respectively.
By construction, the set of $d_L \times d_R$ states $\{ \ket{a, b} \}$ transform into themselves under the symmetries.
		
By their construction, these states will look like the ground state in the bulk of the cylinder, but in general there is no obvious edge Hamiltonian for which they are exact ground states. However, this ambiguity is irrelevant, since the topological invariant are a property of the bulk.

\subsection{Step 2: extracting 1D-SPT order from the ansatz wavefunction}

	Using the CPS trick (App.~\ref{app:CPSrep}), it is trivial to measure the representation $g \to g\indices{^{a' b'}_{a b}}$ acting on the small set of states $\{ \ket{a, b} \}$ (we use raised / lower indices to denote rows / columns of the representation).
To extract the 1D-SPT from the matrices $\{ g \}$, consider first an `on-site' symmetry which does not exchange the edges.
For a long cylinder, the representation will take the form of a tensor product over the left / right edges:
\begin{align}
g\indices{^{a' b'}_{a b}} =  (g^L)\indices{^{a'}_a} (g^R)\indices{^{b'}_b} + \mathcal{O}(e^{-L / \xi})
\end{align}	
The tensor-product decomposition can be found via SVD of $g$ with respect to the left -right decomposition $(a a') \times (b b')$; the SVD should have a unique large singular value.

There is $\textrm{U}(1)$ ambiguity in each $g^L$, which in general form a projective representation of $G$ - this projective representation encodes the 1D-SPT order, as is well documented elsewhere.\cite{PollmannTurnerBO2010, FidkowskiKitaev2011}

For a symmetry $\mathcal{I}$ which exchange the edges, the tensor-product decomposition is altered:
\begin{align}
\mathcal{I}\indices{^{a' b'}_{a b}} =  (\mathcal{I}^L)\indices{^{a'}_b} ({\mathcal{I}}^R)\indices{^{b'}_a} + \mathcal{O}(e^{-L / \xi})
\end{align}
The inversion invariant is detected from $\mathcal{I}^L (\mathcal{I}^L)^T = \pm 1$.\cite{PollmannTurnerBO2010}
More general inversion invariants can be found from the combined representation of inversion and onsite symmetries $\{ {\mathcal{I}}^L, g^{L} \}$. \cite{ChenGuWen2011}

\end{document}